\documentclass[fleqn,usenatbib]{mnras}

\usepackage{newtxtext,newtxmath}

\usepackage[T1]{fontenc}

\DeclareRobustCommand{\VAN}[3]{#2}
\let\VANthebibliography\thebibliography
\def\thebibliography{\DeclareRobustCommand{\VAN}[3]{##3}\VANthebibliography}

\usepackage{graphicx}	

\title[The origin of Alkali-enriched Hot Jupiters]{Super stellar abundances of alkali metals suggest significant migration for Hot Jupiters}

\author[T.O. Hands et al.]{
Tom O. Hands,\thanks{E-mail: tomhands@physik.uzh.ch (TOH)}
R. Helled,
\\
\\Institut f\"ur Computergest\"utzte Wissenschaften, Universit\"at Z\"urich, Winterthurerstrasse 190, 8057 Z\"urich, Switzerland
}

\date{Accepted XXX. Received YYY; in original form ZZZ}

\pubyear{2021}

\begin{document}
\label{firstpage}
\pagerange{\pageref{firstpage}--\pageref{lastpage}}
\maketitle

\begin{abstract}
We investigate the origin of the measured over-abundance of alkali metals in the atmospheres of hot gas giants, relative to both their host stars and their atmospheric water abundances. We show that formation exterior to the water snow line followed by inward disc-driven migration results in excess accretion of oxygen-poor, refractory-rich material from within the snow-line. This naturally leads to enrichment of alkali metals in the planetary atmosphere relative to the bulk composition of its host star but relative abundances of water that are similar to the stellar host. These relative abundances cannot be explained by {\it in situ} formation which places the refractory elements in the planetary deep  interior  rather than the atmosphere. We therefore suggest that the measured compositions of the atmospheres of hot Jupiters are consistent with significant migration for at least a subset of hot gas giants. Our model makes robust predictions about atmospheric composition that can be confirmed with future data from JWST and Ariel. 
\end{abstract}




\section{Introduction}
The formation of Hot Jupiters has been hotly contested for as long as observations of these planets have existed. Giant planets are thought to form either through core accretion or disc instability \citep[see e.g.,][for a review]{Helled2014}, in addition to the actual formation mechanism another key question is whether the observed hot Jupiters  form {\it in situ}  (i.e., at the location of their final orbit) or in much longer-period orbits followed by migration. {\it In situ} formation presents several difficulties regarding the availability of solid material to form a heavy-element core \citep[see e.g.,][]{Rafikov2006}, though \cite{Batygin2016} suggest that {\it in situ} formation is possible if a suitably large core can find its way to the inner disc. This core of roughly 10-20 $M_\mathrm{\oplus}$ must form from solids in the outer part of the disc, and then undergo runaway gas accretion as in the classical core-accretion model, assuming the envelope can cool fast enough \citep{Pollack1996}. In this most of the heavy elements are concentrated in the planetary deep interior with an envelope (and therefore atmosphere) consisting of gas with similar composition to that in the host star, unless substantial mixing takes place.

Alternatively, the giant planets could form at large radial distances and then migrate inward if  they lose orbital energy and angular momentum  after their formation. 
Several mechanisms have been suggested to achieve this, including Kozai-mechanism based migration, planet-planet scattering followed by tidal circularisation, and disc-driven migration \citep[see e.g.,][for a review]{Dawson2018}. On this journey, the planet can accrete planetesimals, which can mix into its atmosphere \citep{Valletta2019} and enrich it in elements that would otherwise be concentrated in the planetary deep interior.  
It has thus been suggested by several authors \citep[see e.g.][]{Oberg2011,Madhu2014,Brewer2017} that over/under-abundances of certain elements in giant planet atmospheres can reveal information about the formation and migration histories of hot Jupiters and help constrain their formation and migration pathways.

The atmospheric chemical composition of giant planets reveals key information on the planetary formation and evolution \citep[see e.g.,][]{Helled2021}. 
Thanks to improved observation techniques, significant progress in such  measurements has been archived in the past decade. 
Various observations imply that hot Jupiters have super-solar atmospheric metallicities. For example, \cite{Nikolov2018} presented evidence of a highly super-solar sodium abundance in hot Saturn WASP-96b. 
\cite{Madhu2014obs} found HD 209458b to have a highly sub-solar water abundance. Similarly, a retrieval performed on spectroscopic data from 10 hot Jupiters by \cite{Barstow2017} found water abundances consistent with anywhere between 0.01 and 1 $\times$ Solar expectations, whilst and \cite{Pinhas2019} retrieved water abundances ranging from small fractions of solar expectations all the way to slightly super-Solar. Other observational constraints include \cite{Sedaghati2017}, who reported abundances of Sodium and water in WASP-19b that are consistent with solar or sub-solar values, and TiO abundances that are notably sub-solar.

In a similar vein, a homogeneously retrieved sample of atmospheric abundances presented by \cite{Welbanks2019} identified a population of hot, giant planets that
\begin{itemize}
    \item have an abundance of H$_2$O that is less than or consistent with expectations based upon their host stars 
    \item have marked over-abundances of the alkali metals Na and K relative to their host stars
    \item display similar levels of enrichment in both Na and K
\end{itemize}
Similar measurements have been reported previously in the planet WASP-127b \citep{Chen2018} -- a super-Neptune in a 4.2 day orbit around its stellar host.
These observations require an enrichment process that provides huge amounts of Na and K whilst simultaneously providing very little H$_2$O/oxygen. On the contrary, certain hot Jupiters present marked deficits in the abundance of the alkali metals relative to the Sun \citep{Pinhas2019}, which implies that the delivery mechanism that provides excess alkali metals to hot Jupiters is not universal. Further measurements of the alkali metal abundances in the host stars of these planets  would help to confirm what this means for planet formation models.

On the theoretical side, various modelling studies have predicted that giant planet atmospheres should be metal enriched relative to their host stars as a result of accreting planetsimals, with sub-stellar C/O ratios as a result of the planetesimals containing little carbon but vast quantities of oxygen \citep{Thorngren2016,Mordasini2016,Brewer2017,Espinoza2017}. 
Whilst all of these models predict an enrichment in refractory elements, none of them explain the high refractory-to-volatile ratio implied by observations. Careful dynamical modelling of the post-formation solid accretion of hot Jupiters is required to understand the exact content of their final atmospheres.

It was recently shown by \cite{Shibata2019} that the overall heavy-element enrichment in hot Jupiters can be explained by disc driven migration. They modelled Jupiters form at 10s of au in a primordial disc, and then accrete planetesimals as they slowly fall toward their host star. The heavy elements within these planetesimals are then mixed into the planet's atmosphere, enriching it with species that otherwise would only be present in the deep interior. 
At first glance, planetesimal accretion during migration appears challenging -- the slow Type II migration that Jupiters undergo in protoplanetary discs is conducive to forming mean-motion resonances (MMRs) between migrating giant planets and planetesimals interior to their orbits. These MMRs would allow the migrating giant planet to rob the interior planetesimals of their angular momentum and shrink their orbits in tandem with the planet, keeping the planetesimals well away from the young planet. However, \cite{Shibata2019} showed that the so called over-stable libration mechanism, described analytically by \cite{Goldreich2014} and seen in numerical simulations by \cite{Hands2018} -  can break the resonances formed during migration, and allow the planetesimals to be accreted. The mechanism results from the interaction of the eccentricity damping from the gas disc and the eccentricity growth that the planet induces upon the planetesimal. There are of course, other mechanisms that could break down or even stop the formation of some of these mean-motion resonances -- stochastic forcing as a result of disc turbulence \citep{Rein2009,Paardekooper2013,Hands2014}, or an external companion perturbing the resonant pair \citep{Hands2016}. There is therefore a complex interplay between disc, planet and planetesimal properties that determines which planetesimals can be accreted, from which points in the disc they originate and therefore what their composition is. The signature of this process is thus present in the final planet's atmosphere as the ratios of different species that condense out at varying points in the disc.

The aforementioned previous results show that migration can lead to heavy element enrichment in giant planets relative to their host stars, but not how the relative proportions of each heavy element are affected by this process. In this paper, we will present the first numerical results that demonstrate that disc-driven migration leads to not only a general over-abundance of heavy elements in hot gas giant atmospheres, but also creates a preferential overabundance of refractory elements relative to water. Given that {\it in situ} formation places these heavy elements only in the core of the planet, we believe this trait to be the signature of migration in the history of a gas giant.

\section{Can accretion during migration preferentially deliver refractory-rich, volatile-poor material?}
\subsection{The migration mechanism}
In-situ formation of hot Jupiters limits the accretion of a forming giant planet, since the lack of radial movement means the planet can only accrete from the area local to it in the disc and therefore never has the opportunity to accrete planetesimals from elsewhere in the disc, unless they are thrown into the innermost part of the system by some mechanism. 
As mentioned above, one would expect the refractory material in a giant planet to be in the deep interior rather than the atmosphere in this case, unless significant core erosion takes place, which would not explain the observations of \cite{Welbanks2019}. Given this fact and following the promising results of \cite{Shibata2019}, we choose to concentrate our efforts on short-period gas giants that reach their final location by disc-driven migration rather than dynamical methods such as tidal migration \citep[e.g.,][]{Dawson2018}, but also perform a subset of simulations without gas drag. For the vast majority of our simulations we consider planetesimals of 10km radius, but also run a subset with 1km radius for comparison. The majority of giant planets are expected to undergo migration in the so-called "Type-II" regime \citep[see e.g.,][for a review]{Baruteau2014}, whereby the planet is sufficiently massive to open a considerable gap in the disc which is largely devoid of gas and therefore exerts little torque on the planet. This regime is slow relative to other forms of planet migration, and was traditionally thought to lead to migration time-scales equivalent to the viscous time-scale of the host disc, however,  more recent studies suggest the migration rate might be independent of the viscous behaviour of the disc \citep{Durmann2015} or correlated with, but not equal to the viscous time-scale \citep{Robert2018}.

Depending upon disc conditions and planetary mass, some of the planets studied by \cite{Welbanks2019} might not form a gap deep enough to push them into the Type II regime. With only a partial gap forming in the disc, they might instead enter the Type III regime, also known as runaway migration. Canonically the migration time-scale in the type III regime is several tens of orbits \citep[see e.g.,][]{Nelson2018} -- much faster than the type II regime. In this case we might reasonably expect very few mean-motion resonances to be formed between the migrating giant planet and planetesimals interior to it, in turn allowing the planet to accrete material without the assistance of the "overstable-librations" mechanism. Thus, it may be easier to enrich Saturn-mass planets than their Jupiter mass brethren, and we may see less of a dependence on planetesimal size compared to the results of \cite{Shibata2019}. For the rest of this paper, we will investigate how the migration rate and other disc parameters affect the locations from which planets accrete planetesimals, and in turn how this alters the eventual composition of these planets.

\subsection{Estimating the mass of solids required for enrichment}
\begin{figure}
\begin{center}
\includegraphics[width=0.99\linewidth]{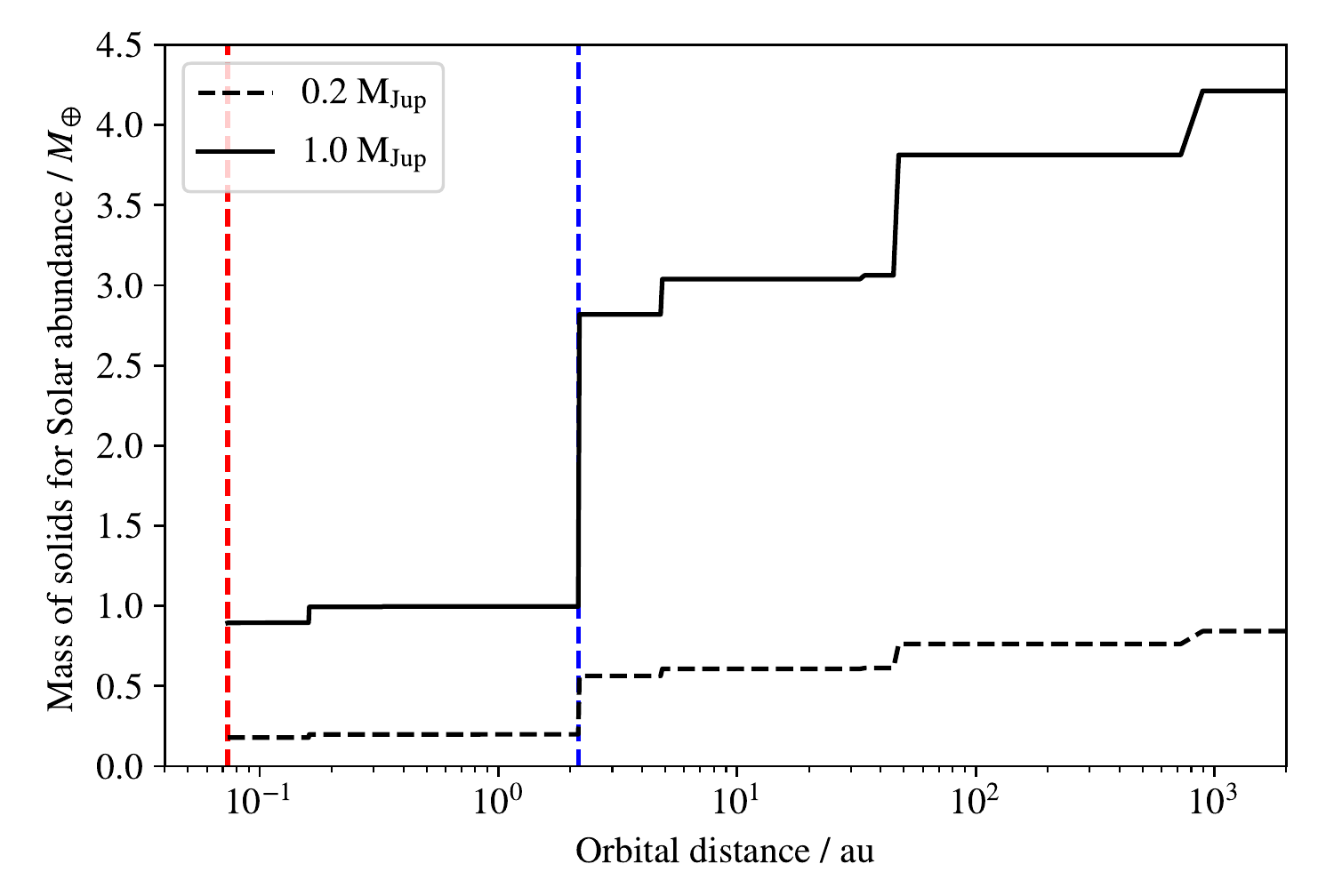}

\caption{An estimate of the mass of solid material that a 0.2 $M_\mathrm{jup}$ and 1 $M_\mathrm{jup}$ planet that formed with a pure H/He atmosphere need to accrete from a given orbital radius to achieve stellar abundances of the local solids. We assume a disc with a temperature profile $T(R) = T_{1\mathrm{au}} / R^{1/2}$ with $T_{1\mathrm{au}} = 268K$ -- equivalent to our numerical disk model. The red vertical line is the 50\% condensation temperature of K (993K) and the blue line is the water ice-line (183K). The estimates are based on the assumption that a given element condenses out entirely from the gas at solar abundance at $T_{50}$ and are designed only as a guide. Material accreted from within the iceline at 2.1au preferentially enriches the atmosphere in alkali metals without providing a similar enrichment in water. This means for instance, that a Jupiter mass planet can acquire a 10x enrichment by accreting a relatively low 10 $M_\oplus$ of material within the iceline, whilst a 0.2 $M_\mathrm{jup}$  mass planet requires only 2 $M_\oplus$ . \label{fig:macc} Temperature data from \protect\cite{Wood2019}.}
  \end{center}
\end{figure}
We consider a scenario in which the planet migrates and planetesimals enter its gaseous envelope. In this case, the planetesimals may fragment or ablate, preferentially enriching the inner or outer atmosphere respectively with their contents \citep{Valletta2019}. In order to produce a giant planet with an envelope enriched in alkali metals but little water, the giant must accrete solids from the inner disc at a location where alkali metals condense, whilst simultaneously avoiding planetesimals formed in the ice-rich regions beyond the snowline. Of course, accreting material from within the water ice-line also leads to over abundances of other metals with high condensation temperatures such as Fe, Mg and Li, and not just the Na and K that have been measured. For Jupiter to have Solar metallicity (taken here to be $Z = 0.014$), it would need to contain around 4.5 $M_\oplus$ of heavy elements, while a 0.2 $M_\mathrm{Jup}$ mass planet requires 0.9 $M_\oplus$. Assuming this quantity was accreted entirely as solids (i.e. the initial envelope of the planet consisted only of H and He gas), at least 4.5 $M_\oplus$ of planetesimals formed exterior to the ice line would be required to enrich the Jupiter -- the actual number being dependent upon the accretion efficiency. Within the water ice-line, the lack of water ice reduces the mass of available solids by as much as 75\% \citep[see e.g.][]{Hayashi1981}. As a result, achieving a solar composition of the elements within this region with a dearth of oxygen would require significantly less accretion - 1.13 and 0.23 $M_\oplus$ for a Jupiter and a 0.2 $M_\mathrm{Jup}$ mass planet respectively

We can make a more accurate estimate of the quantity of solids required to produce a given enrichment of alkali metals by considering a Solar-composition protoplanetary disc and using pre-computed 50\% condensation temperatures ($T_{50}$). This is the temperature at which 50\% of the total abundance of a given element would condense out of a protoplanetary disc \citep{Wood2019}. Then, for a given temperature $T(R)$ -- which can be translated to a radial location in the disc for a specific disc model -- we can find the elements for which $T_{50} > T(R)$ and by using the solar abundances, compute an approximation of the elemental composition of the solids at this disc location. This method implicitly assumes that all elements for which $T_{50}$ have  condensed out in solar proportions together, which can lead to up to a 2x error in our assumed compositions at a given radius and miss certain elements that are less than 50\% condensed out. However, given the errors in the atmospheric retrievals and the other modelling techniques we use later in this paper, we are confident that this  uncertainty does not affect our conclusions.

For condensation temperatures and solar abundances we use the machine-readable tables given by \cite{Wood2019} and \cite{Wang2019}, respectively. Figure \ref{fig:macc} displays an example calculation for a disc profile equivalent to our fiducial one (discussed in detail in the next section). Note that the amount of mass required for a given level of alkali metal enrichment is relatively flat within the water ice-line since most metals condense already at very high temperatures close to the star. An important distinction between this approach and the simplified approach based solely on stellar metallicity above relates to the aforementioned canonical 1:4 difference in solid mass across the water ice line.  In reality at distances of 1s or 10s of au, much of the carbon and nitrogen in a protoplanetary disc has not condensed out, and thus these water rich regions contain only around 3x more mass in solids than regions within the water ice-line. In figure \ref{fig:macc} we have to look to 100s of au before we find a region where 4x the amount of solids have condensed out relative to inside the ice line.

In this paper, for simplicity we assume that only solids from outside the water ice-line which contribute to the water content of a planet's envelope. As a result, our analysis corresponds only to the abundance of water in the envelope, rather than the overall abundance of oxygen. 
In addition, $\sim$20\% of oxygen could condense out into rock at temperatures higher than the condensation temperature of water \citep{Lodders2003}. Therefore, our estimates for the water abundances should be taken as lower bounds for oxygen abundance.

\section{Method}
With an understanding of where in a disc a giant planet should accrete solids from to achieve a given overabundance, we use a bespoke $N-$body code to model the interaction of migrating proto-planets with planetesimals to investigate the accretion history of a migrating giant planet. 
This code uses the reversible, GPU-based time-stepping method described in \cite{Hands2019}, coupled with a GPU implementation of  the '4a' symplectic integrator described by \cite{Chin2005} and the migration and eccentricity damping prescriptions described in \cite{Hands2014}. Our planets begin fully formed and we do not consider any gas accretion during migration. In addition, the initial atmospheric is assumed to contain no oxygen (see below for further discussion). In our fiducial simulations, we initialise giant planet at 5au around a solar mass star with zero eccentricity and inclination, and with masses of 0.2, 0.4 and 1.0 $M_{\mathrm{jup}}$ in order to cover the range of observations by \cite{Welbanks2019}. 

Throughout this work we vary the migration time-scale $\tau_\mathrm{a}$ between $1 \times 10^3$ and $1 \times 10^5$ yr of the planets to explore the possible parameter space between extremely fast type III migration and type II migration on a roughly viscous time-scale \citep[see e.g.][]{Nelson2018,Robert2018}. We halt simulations once the planets reach 0.25au due to computational constraints, though we have confirmed that the results are not sensitive to stopping migration closer to the star. We damp the planet eccentricity with dimensionless damping parameter $K = 10$. This parameter defines the factor by which eccentricity damping is faster than the migration time-scale (see \cite{TanakaandWard2004,Hands2014} for more information). We do not explore different values of $K$ since there are no other massive bodies to perturb the orbit of the giant. We treat planetesimals as test particles that feel only the gravitational pull of the star and planet, and spread 5000 of them them with a surface density that varies with $1/r$ throughout the region $R=0.25$au to $R=4.5$au in cases where the giant planet begins at 5au.  We initialise the eccentricity ($e$) and inclination ($i$) of each planetesimal according to a Rayleigh distribution\footnote{We note that we expect the gas drag to quickly reduce the initial inclinations and eccentricities of the planetesimals to zero in simulations with drag enabled.}.

Additionally, we have added a GPU-based implementation of gas drag on planetesimals, where the gas drag on an individual planetesimal is computed as
\begin{equation}
    F_D = \frac{1}{2} \rho v^2 C_D A, 
\end{equation}
where $\rho$ is the gas density, $v$ is the differential velocity between the planetesimal and the gas, $C_D$ is the drag coefficient, and $A$ is the cross sectional area of a planetesimal. $A$ we take to be the area of a circle with the same radius as a planetesimal, and in most our simulations we take the radius of a planetesimal to be 10km.
Computation of the gas density $\rho$ at any given point requires the full definition of a disc model.
The numerical results of \cite{Li2015} suggest that the median disc mass around such a star in the class II phase would be of order 0.2 stellar masses, and we take this as the initial mass of our disc after formation. Assuming that the disc surface profile varies as $\Sigma_{1au}/r$ and the radial extent of the disc is from 0.1 to 50au, then the surface density of this disc at 1 au is  $\Sigma_{1au} = 6.37 \times 10 ^{-4} M_\odot/\mathrm{au}^2$. However, since the disc evolves considerably between the time from its formation to the formation of a giant planet, we must account for the loss of mass from the disc. We model the mass loss as an exponential \citep{Mamajek2009} with e-folding time $8 \times 10^5$ yr and consider the disc to be $3 \times 10^6$ yr old, consistent with estimates of several Myr for how long Jupiter would take to reach runaway growth in the Solar system \citep[see e.g.,][]{Pollack1996,Lissauer2009,Monga2015,Kruijer2017}. As a result of this, we start our simulations with $\Sigma_{1au} = 1.5 \times 10 ^{-5} M_\odot/\mathrm{au}^2$. 
The  density at a given point can be computed from the surface density and the scale height $H$ as 
\begin{equation}
\rho(R, z) = \rho_0 \mathrm{exp}\left(\frac{-z^2}{2H^2}\right),   
\end{equation}
where $H$ is the disc scale-height at radial location $R$ and the mid-plane density $\rho_0 = \Sigma/(\sqrt{2 \pi}H)$. We use a moderately flaring power-law to set the scale-height of the disc $H(R) = H_{1\mathrm{au}} (R/1\mathrm{au})^{(5/4)}$, such that the mid-plane temperature varies as $T(R) \propto R^{-1/2}$ \citep[found to be a reasonable profile by e.g.,][]{Dalessio1998}. This temperature profile does not vary in time, since we would expect minimal change in the disc temperature profile over the majority of the time-scales considered here. We set $H_{1\mathrm{au}} = 0.033 \mathrm{au}$ which places the water ice-line of 183K at around 2.1au. We consider a planetesimal to have been captured by the planet if it enters the planet's atmosphere or if the planetesimal spends 100 yr continuously in an $e < 1$ orbit relative to the planet with an apocentre that is less than 0.5$r_H$, where $r_H$ is the planet's Hill radius.

Finally, we estimate the quantity of solids available for giant planets to accrete in such a disc model, assuming solar metallicity $Z = 0.014$ and a solid surface density profile that follows that of the gas. We assume based on our model in figure \ref{fig:macc} that only a quarter of the heavy elements are condensed out of the disc within the ice-line ($a < 2.1$au) since a significant amount of the oxygen in the disc does not condense out until solid water can form, whilst the full fraction of 0.014 of the disc's heavy element mass is condensed out exterior to this. This is consistent with \cite{Hayashi1981}. Integrating across these two portions of the disc suggests that 54 $M_\oplus$ of solids are available for accretion to a migrating giant. 

\subsection{Analysis}
We do not consider our test particles to be 1:1 representations of the planetesimals in a protoplanetary disc -- indeed, representing a population of 50 $M_\oplus$ of 10km radius planetesimals would be prohibitively computationally expensive, requiring over $3 \times 10^{10}$ particles. 
Instead, we consider them to be a tracer population, and therefore bin the planetesimals according to their initial semi-major axes and compute the fraction captured from each of these initial bins during the simulation. 
This allows us to impose whatever surface density features we desire during analysis. Using the initial $1/r$ surface-density power-law for the planetesimals  ensures there is enough resolution in each bin to achieve statistically significant results, since a uniform surface density places relatively few planetesimals in the inner disc. Nontheless, for our analysis, we consider an initial planetesimal surface number density that indeed scales as 1/r and therefore traces our aforementioned gas surface density, but with a 3x jump at the ice-line to represent the additional freezing out of water into solids. We consider this initial population to represent 50 $M_\oplus$ of rocky material for most of our analysis based on estimate of planetesimal mass in the previous section. We define the enrichment factor, $E$, to be the relative overabundance of either ice-poor/refractory-rich ($E_\mathrm{ref}$) or ice-rich ($E_\mathrm{vol}$) material accreted, compared to the amount that would be expected for a planet with solar composition. Therefore we require models with $E_\mathrm{vol} \approx 1$ and $E_\mathrm{ref} > 1$ to explain the observations. The enrichment factors are computed as
$$E_\mathrm{vol} = \frac{M_\mathrm{vol}}{M_\mathrm{vol,solar}}$$
and
$$E_\mathrm{ref}= \frac{M_\mathrm{vol}}{M_\mathrm{vol,solar}}  + \frac{M_\mathrm{ref}}{M_\mathrm{ref,solar}} $$
where $M_\mathrm{vol}$ and $M_\mathrm{ref}$ are the masses of water-rich/volatile and water-poor/refractory material accreted, respectively, and $M_\mathrm{vol,solar}$ and $M_\mathrm{vol,solar}$ are the masses required for each planet to have a solar composition of each as computed in section 3. 
Note that since material beyond the water ice-line also includes solar composition quantities of refractory elements, the mass of material accreted from outside the ice-line is also required for the computation of $E_\mathrm{ref}$.  It must also be noted that the water-poor material changes the abundance of many species and not just the Na and K that are observationally measured, and therefore our results have implications for the abundances of many other elements within the atmospheres of hot giant exoplanets. Finally, our calculations of $E_\mathrm{vol}$ do not account for volatiles such as CO and CO$_2$ that are present in gaseous form outside the water ice-line and therefore incorporated into the planet's envelope during runaway accretion. 
From here on we will refer to $E_\mathrm{ref}$ and alkali metal enrichment interchangeably, and the same with $E_\mathrm{vol}$ and water enrichment.
\section{Results}
\begin{table*} 
\begin{tabular}{|l|l|l|l|l|l|l|l|l|l|}
\hline
$M_p (M_\mathrm{jup})$ & $\tau_a (\mathrm{yr})$ & Drag? & $a_p$ (au) & $r_\mathrm{plan}$ (km) & $M_\mathrm{plan} (M_\oplus)$  & $m_\mathrm{ref} (M_\oplus)$ & $m_\mathrm{vol}(M_\oplus)$ & $E_\mathrm{ref}$ & $E_\mathrm{vol}$ \\ \hline
0.2                    &     $1 \times 10 ^3$   & n     & 5        &        10           &           50        &   0.414 & 0.713 & 3.579 & 1.265\\ \hline
0.2                    &     $1 \times 10 ^4$    & n     & 5        &        10           &         50          &  0.129 & 0.516 & 1.636 & 0.915\\ \hline
0.2                    &    $1 \times 10 ^5$     & n     &   5         &      10             &      50             &  0.0 & 0.336 & 0.596 & 0.596\\ \hline
0.2 & $1 \times 10 ^3$ & y & 5 & 10 & 50 & 0.896 & 0.843 & 6.507 & 1.496\\ \hline 
0.2 & $1 \times 10 ^4$ & y & 5 & 10 & 50 & 1.728 & 0.919 & 11.299 & 1.631\\ \hline 
0.2 & $1 \times 10 ^5$ & y & 5 & 10 & 50 & 0.853 & 5.417 & 14.382 & 9.611\\ \hline 
0.2 & $1 \times 10 ^4$ & y & 10 & 10 & 50 & 1.604 & 1.415 & 11.482 & 2.511\\ \hline
0.2 & $1 \times 10 ^4$ & y ($\Sigma_0 \times 2$) & 5 & 10 & 50 & 1.384 & 1.148 & 9.78 & 2.037\\ \hline
0.2 & $1 \times 10 ^3$ & y & 5 & 1 & 50 & 0.804 & 0.738 & 5.807 & 1.309\\ \hline 
0.2 & $1 \times 10 ^4$ & y & 5 & 1 & 50 & 1.22 & 2.781 & 11.761 & 4.934\\ \hline 
0.2 & $1 \times 10 ^5$ & y & 5 & 1& 50 & 0.11 & 3.61 & 7.019 & 6.405\\ \hline 

0.4 & $1 \times 10 ^3$ & y & 5 & 10 & 50 & 0.937 & 0.908 & 3.426 & 0.805\\ \hline
0.4 & $1 \times 10 ^4$ & y & 5 & 10 & 50 & 2.554 & 2.015 & 8.932 & 1.788\\ \hline
0.4 & $1 \times 10 ^5$ & y & 5 & 10 & 50 & 1.031 & 5.728 & 7.966 & 5.081\\ \hline


1.0 & $1 \times 10 ^3$ & y & 5 & 10 & 50 & 1.647 & 1.163 & 2.255 & 0.413\\ \hline
1.0 & $1 \times 10 ^4$ & y & 5 & 10 & 50 & 2.135 & 5.026 & 4.173 & 1.783\\ \hline
1.0 & $1 \times 10 ^5$ & y & 5 & 10 & 50 & 0.406 & 6.894 & 2.9 & 2.446\\ \hline

1.14 & $1 \times 10 ^3$ & y & 5 & 10 & 50 & 1.814 & 1.134 & 2.134 & 0.353\\ \hline
1.14 & $1 \times 10 ^4$ & y & 5 & 10 & 50 & 2.2 & 5.081 & 3.741 & 1.582\\ \hline
1.14 & $1 \times 10 ^5$ & y & 5 & 10 & 50 & 0.49 & 7.226 & 2.73 & 2.249\\ \hline
1.14 & $1 \times 10 ^5$ & y & 3 & 10 & 50 & 0.464 & 6.28 & 2.41 & 1.954\\ \hline
\end{tabular}
\caption{Results of our fiducial analysis for all planet masses and migration time-scales. $r_\mathrm{plan}$ is the radius of the planetesimals in a given simulation used to compute gas drag.  $M_\mathrm{plan}$ is the initial total mass of planetesimals distributed interior to the giant, an assumption which we change during analysis.\label{tab:one} }
\end{table*}

\begin{figure}
\begin{center}
\includegraphics[width=0.99\linewidth]{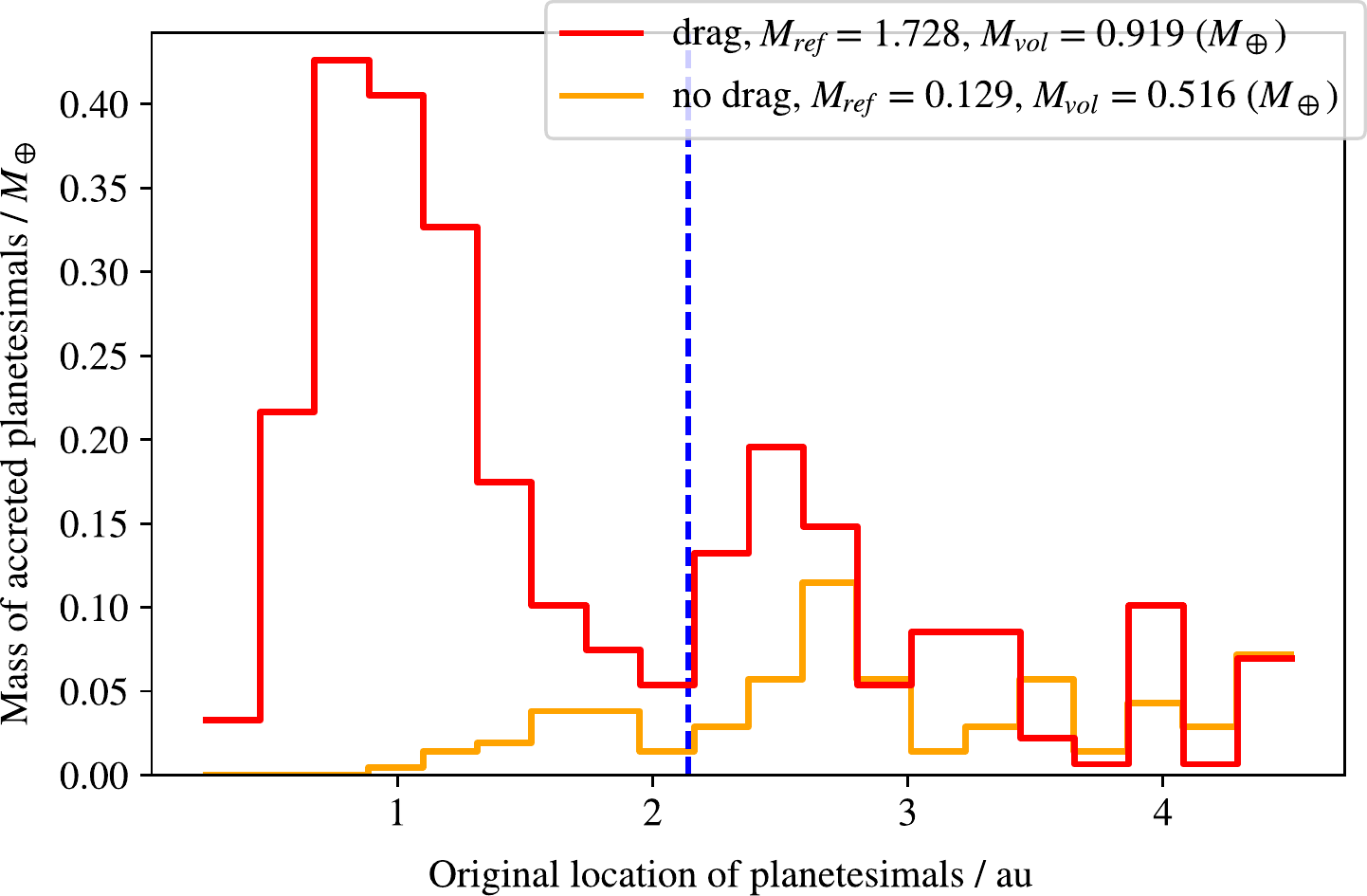}
\caption{Initial locations within the disc of planetesimals accreted by a 0.2 $M_\mathrm{Jup}$ planet migrating with a 10,000yr time-scale both with and without gas drag acting on the planetesimals. The water ice-line is shown in blue. The captured mass of planetesimals is far higher with gas drag enabled, due to resonance breaking cause by the overstable librations mechanism first identified by \protect\cite{Goldreich2014}. This also allows the planet to accrete preferentially from within the water ice-line, providing an enrichment of alkali metals but only the expected solar abundance of water. \label{fig:fig1} }
  \end{center}
\end{figure}

\begin{figure}
\begin{center}
\includegraphics[width=0.99\linewidth]{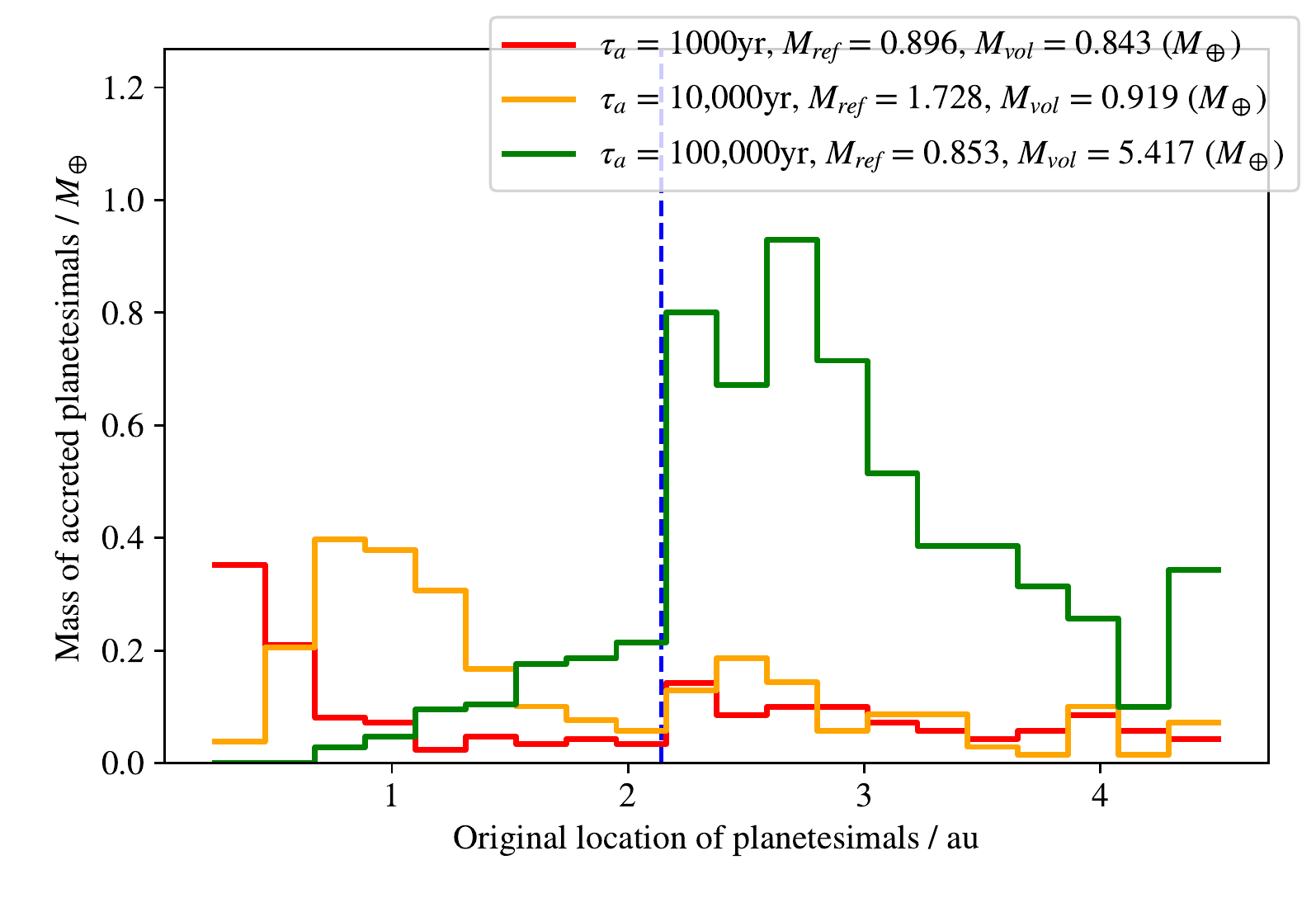}

\caption{Accretion locations for a 0.2 $M_\mathrm{Jup}$ planet migrating from 5.2au with gas drag enabled. The 1,000 and 10,000 yr migration time-scales deliver a roughly solar abundance of water-rich material from outside the ice line, with both models leading to an enrichment of refractory-rich material but the 10,000 yr time-scale leading to almost twice the enrichment of its faster counterpart. The 100,000 year migration time-scale allows the planet to accrete from outside the water ice line, leading to an almost 10x enrichment in water compared to what would be expected from solar composition. }\label{fig:fig2}
  \end{center}
\end{figure}

Our results are summarised in table \ref{tab:one}. Specific examples of where 0.2 $M_\mathrm{Jup}$  mass planets accrete solids from in our runs are show in figure \ref{fig:fig1} for varying gas drag and figure \ref{fig:fig2} for varying migration time-scales. Simulations with aerodynamic drag allow planets to accrete more material overall, and shorter migration time-scales drive clear preferential accretion of material from within the ice-line, leading to $E_\mathrm{ref} > 1$. In \emph{all} cases with gas drag, we find $E_\mathrm{ref} > 1$ and $E_\mathrm{ref} > E_\mathrm{vol}$, even when $E_\mathrm{vol} > 1$. This trend to be in good agreement with the observations, which almost universally predict higher relative abundances of alkali metals compared to water. Longer migration time-scales or a lack of gas drag skew the accretion more toward the water-rich solids exterior to the ice-line, which immediately points to Na and K enriched giants being formed in more massive discs with larger surface densities and therefore increased gas drag and faster planet migration. Given this trend, we chose not to run more simulations without gas drag and focus on the more realistic simulations including drag.

From table \ref{tab:one} and figure \ref{fig:fig2}, it is clear that for a sub-Saturn-mass object, a migration time-scale of 10,000yr is preferred for generating high levels of alkali metal enrichment. 
This migration time-scale gives the planet a water abundance similar to solar composition, whilst the alkali metals are 10 times more abundant. Of course, these numbers scale linearly with the mass of planetesimals available to the planet. The shorter time-scale of 1000yr still generates preferentially higher abundances of alkali metals, though not to the same level -  enriching the alkali metals by a factor of 6 rather than 10. These migration time-scales are close to those expected of Type III migration on a time-scale of tens of orbits for a young sub-Saturn \citep[see e.g.][for a review]{Baruteau2014}. At the longer migration time-scale of 100,000 yr -- representative of such a planet migrating in the type II regime -- the 0.2 $M_\mathrm{Jup}$ planets still accretes several times the expected solar abundance of alkali metals, but the abundance of water is also enriched compared to solar levels by almost an order of magnitude. However, the over-abundance of water still doesn't match the over-abundance of alkali metals.
This is a result of the slow migration time-scale capturing almost all the innermost planetesimals in mean-motion resonances and pushing them inwards with the planet. We therefore posit that gas giants with large over abundances of both water and alkali metals may have been subject to slow migration, possibly as a result of opening large gaps in their parent discs. 
This in turn points to low viscosity and small scale-heights/temperatures in the disc, both of which allow for easier gap opening \citep[see e.g.][]{Crida2006}. With the longer migration time-scale but no gas drag, the planet cannot achieve even solar levels of Na, K or H$_2$O, which suggests that giants with sub-stellar metallicities are formed in discs with very low gas drag and/or migrate by an altogether different mechanism once the gas has dissipated.

\begin{figure*}
\begin{center}
\includegraphics[width=0.99\linewidth]{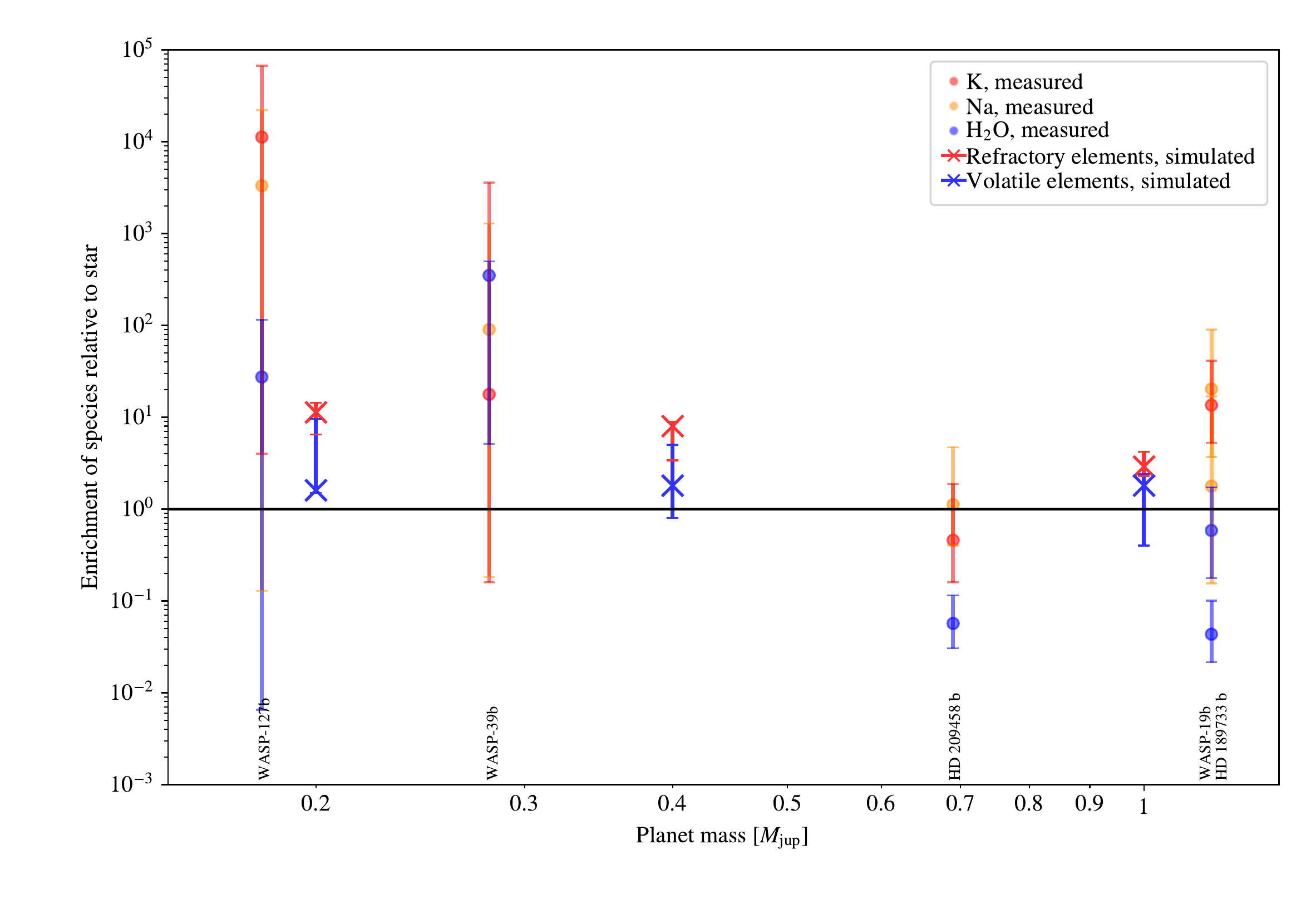}

\caption{Enrichments relative to stellar expectations in water/volatiles (blue), potassium/refractory elements (red) and sodium (yellow), for both exoplanets from \protect\cite{Welbanks2019} (dots) and simulated planets from this work (crosses). Error bars for observational data taken from \protect\cite{Welbanks2019}, while error bars for simulated planets show the range of enrichment achieved by varying the migration time-scale. We exclude observations that do not include abundance measurements for at least one alkali metal and water. Note that the enrichment in the simulations scales linearly with mass of planetesimals in the model disc, the the ratio between refractory and volatile material is maintained when varying planetesimal mass. Data points for WASP-19b and HD 189733b overlap due to their similar masses. \label{fig:fig4} }
  \end{center}
\end{figure*}

Our higher mass (0.4 $M_\mathrm{Jup}$ and 1.0 $M_\mathrm{Jup}$) simulations display a similar trend, with slower migration allowing for the accretion of more water and bringing $E_\mathrm{vol}$ nearer to $E_\mathrm{ref}$ while maintaining the important $E_\mathrm{ref} > E_\mathrm{vol}$. The overall masses accreted by these planets are greater but note that this does not lead to greater enrichment relative to the star. Figure \ref{fig:fig4} shows the results of these simulations in context with each other and the observational results from \citep{Welbanks2019}. Our simulations clearly reproduce the trend observed by \cite{Welbanks2019} where more massive planets are less well enriched in both water and refractory elements. 

In order to check the dependence of our results to the model assumptions, we reran the case of a 0.2$M_\mathrm{Jup}$ planet with 10,000yr migration time-scale with a) a 10au starting location for the giant b) a $2 \times $ higher surface density. The results of these simulations are listed in table \ref{tab:one}. We find that our conclusions are unchanged. Similarly, we reran our three 0.2$M_\mathrm{Jup}$ simulations with our three fiducial migration time-scales but with planetesimals of 1km radius rather than 10km. These results are also in table \ref{tab:one}. The overall trend is similar, with a migration time-scale of 10,000yr generating the highest relative levels of alkali metal enrichment. The total mass accreted in each of these simulations is reduced compared to the 1km radius simulations, though given the complex interplay of mean motion resonances and eccentricity damping from the gas drag, it is plausible that this would be different if we also considered a different gas surface density with the smaller planetesimals. Importantly though, in all these simulations, we still find that the planet accretes significantly super-stellar abundances of alkali metals, without producing equal enrichments in water.

In almost all of our simulations, we find 
that migrating gas giant giants accrete relatively more refractory-rich material than volatile-rich material.  Our results therefore reinforce the conclusions of \cite{Shibata2019} in showing that migration increases the metallicity of giant planets relative to their host stars, but we additionally show that for planets formed close to their host stars, this enrichment is skewed towards refractory elements that condense within the radius of the water ice-line. Our model therefore predicts an equal enrichment in all refractory elements for planets with measured over-abundances of Na and K. Our study explains why \cite{Welbanks2019} find that Na and K are always equally over-abundant in their transiting planet sample, and we predict that further measurements of other metals would yield similar levels of enrichment.

Our results show that hot giant planets with super-stellar alkali metal abundances are consistent with formation beyond the snowline followed by planetary migration. If one considers the possibility of in-situ formation for hot Jupiters in the traditional core-accretion framework (even barring the difficulties of forming planets at such small radii), then all of the refrarctory elements at the formation location are in solids, and the the initial heavy-element core would contain all of these species including Na and K. The gas accreted by the core in the runaway phase would of course contain (almost) solar abundances of oxygen and other volatiles as these are not condensed out of the disc at such high temperatures. This gas would not, however, contain any refractory elements, leading to an atmosphere that meets solar expectations for volatiles but contains almost no refractory material.

\section{Comparison to known exoplanets}
Here we consider the implications of our results for several of the planets with observationally constrained abundances, shown in Figure \ref{fig:fig4}. 
We begin by considering WASP-127b, a 0.18 $M_\mathrm{Jup}$ planet in a 4.2 day orbit. The measured abundances in this planet's atmosphere have large errors, making it difficult to make strong statements and perform a direct comparison to our models, nevertheless, we note that the enrichments generated in all of our $0.2$  $M_\mathrm{Jup}$ models overlap with the measurement uncertainties for this planet. 
There is evidence from \cite{Allart2020} that WASP 127b is on a retrograde orbit, which likely points to migration via dynamical means rather than gas-disc migration. This could help to explain the potentially enormous enrichments in both alkali metals and water, which cannot be explained by our models. Instead, we suggest that WASP-127b may have undergone one or more giant impacts on its route to its present orbit, and intend to study this potential origin for alkali-metal enriched planets in future work.

Second, we consider the planet HD 209458b, a 0.69 $M_\mathrm{Jup}$ planet with a well-constrained ultra-low water abundance. This planet appears to be well aligned with its stellar host's equator \citep{Winn2005}, consistent with disc-driven migration since interactions with the gas disc would damp any dynamically-induced inclination. Recent studies suggest that the detection of sodium in this planet's atmosphere may be a false positive \citep{Casasayas-Barris2020}. However, assuming the values reported by \cite{Welbanks2019} are correct, the water abundance for this planet is not reproduced by our simulations. The planet HD 189733b is also well-aligned with the spin of its host star \citep[see e.g.,][]{Gregory2008} and has a similar water depletion that is not reproduced here. We suggest that our models do not reproduce such low water abundances due to the limited parameter space we explored regarding the disc properties and the primordial atmospheric composition. 
For instance, a steeper radial surface density profile would reduce the availability of volatile rich rocks. 
We intend to consider a wider range of more complex disc models as well as various primordial planetary compositions in future research. Finally, we consider WASP-19b a 1.14 $M_\mathrm{Jup}$, and the planet closest to our models in terms of enrichment fractions and the best constrained from a compositional point of view. WASP-19b the planet is aligned with the spin of its host star \citep{Hellier2011}. The figures from \cite{Welbanks2019} suggest this planet has a 1.72x sodium enrichment relative to its host star but a 0.5x water enrichment. As seen in figure \ref{fig:fig4}, these figures and their errors are consistent with our models of 1 $M_\mathrm{Jup}$ planets formed by migration and including gas drag.

\begin{figure}
\begin{center}
\includegraphics[width=0.99\linewidth]{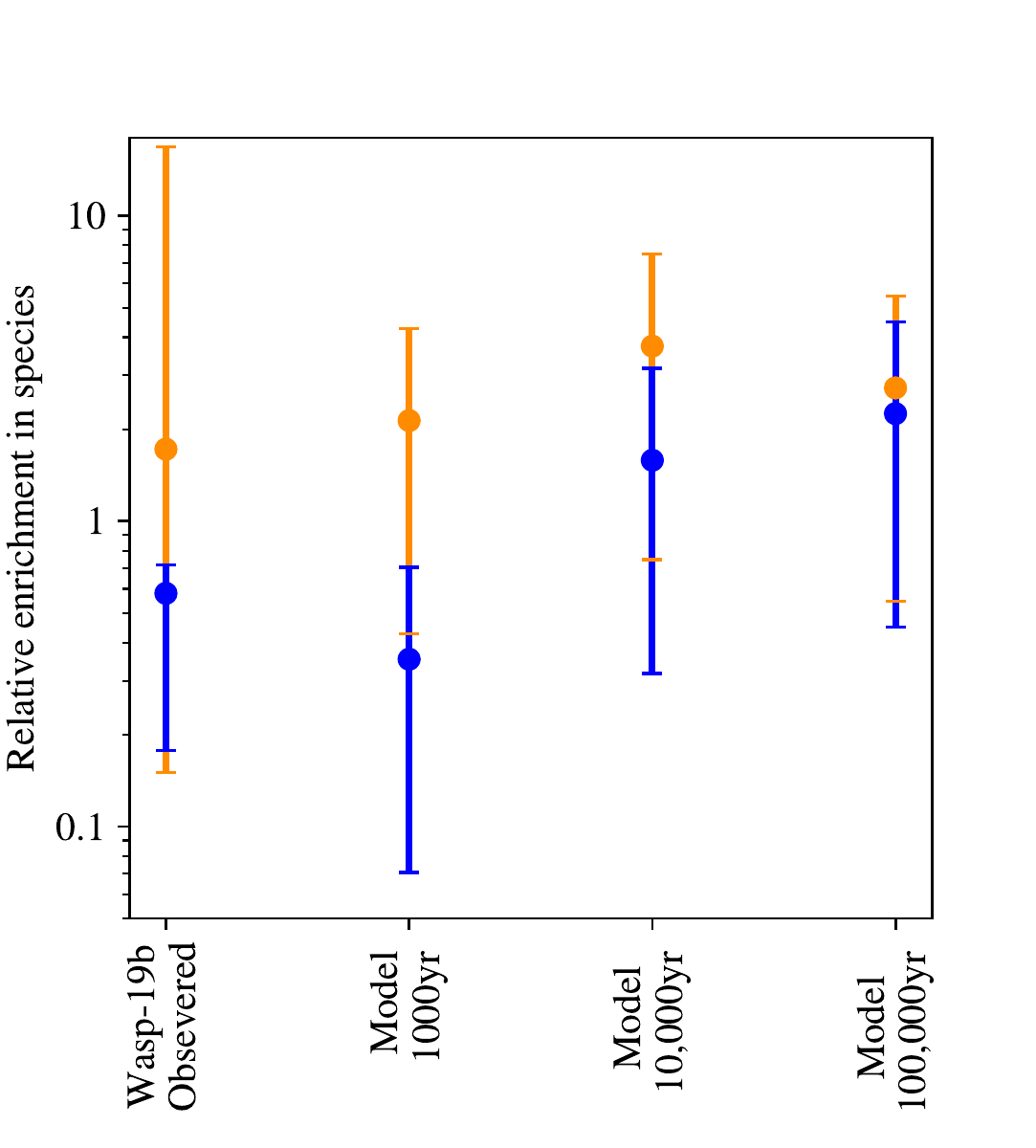}

\caption{Comparison of measured enrichments (refractories/sodium in orange, water in blue) in WASP-19b against models of a 1.14$M_\mathrm{Jup}$ planet migrating from 5au to 0.25au. The error bars on the simulations represent altering the mass of planetesimals available between 10 and 90 $M_\oplus$. The $\tau_\mathrm{mig} = 1000yr$ provides the nominal best fit to the observations, but would represent very fast migration for a super Jupiter. The slowest migration time-scale still produces enrichments consistent with observations with low planetesimal masses.
\label{fig:wasp19} }
  \end{center}
\end{figure}

To explore the apparent close agreement between our models and the composition of WASP-19b, we perform several extra simulations of a 1.14 $M_\mathrm{Jup}$ planet around a 0.968 $M_\odot$ star with gas drag, keeping the rest of our parameters the same \citep[planet parameters from][]{Lendl2013}. The raw results from these simulations are available in table \ref{tab:one}. Despite the good measurements of WASP-19b, the errors are still relatively large - spanning two orders of magnitude in refractory enrichment. Nevertheless, we find that by reducing the mass of planetesimals available to the planet, it is easy to produce planets with enrichments consistent with those derived by \cite{Welbanks2019}. With our fiducial 50 $M_\oplus$ of planetesimals, we find only the 1000yr time-scale reproduces the observed water enrichment. 
However, reducing the planetesimal mass scales the accretion of volatile and refractory material down by the same amount, achieving values consistent with the observed water and Na enrichments, as seen in figure \ref{fig:wasp19}. A better fit could be achieved by modifying the surface density profile of the gas and planetesimals, the scale height of the disc and other parameters, but it is clearly promising that our simulations with fixed disc parameters can produce abundances that are in excellent agreement with the observed ones. 

\section{Discussion}
In this study, we have not attempted to model the long term behaviour of the planetesimals, nor use a more sophisticated model for the locations at which planetesimals form. In principle it is possible to  marry the migration and N-body modelling described here to a model of disc evolution and planetesimal formation \cite[see e.g., that of][]{Drazkowska2018}. This would result in a less uniform initial distribution for planetesimals and may affect the final composition of the planet, but such a model would also have to consider pebble accretion which could further affect the final planetary composition.

While we have studied a range of migration time-scales with the same underlying disc drag model here, it should be noted that the migration time-scale is a product of the exact disc and planet parameters. We intend to use hydrodynamical simulations in future research to self-consistently compute migration tracks for our given disc model. Such simulations could also determine how disc properties such as viscosity and scale height which affect gap opening, can change the locations from which planetesimals are accreted. However, we are confident having covered several orders of magnitude in migration time-scale here, that our results will hold regardless of the exact migration speed. Our models do not reproduce the ultra-low water abundances of HD 189733b and HD 209458b, but we note that we only considered one disc surface density profile and only two initial locations for the planets. Further work should concentrate on ascertaining if physically-realistic disc models and starting locations can reproduce these very low water abundances. 

It should be noted that there is also the possibility of later accretion of small objects which could provide both refractory elements and volatiles to a hot gas giant that is already in place, changing the enrichment levels in the planetary atmosphere. This, however, requires an external companion that can either scatter planetesimals inward or cause Kozai-Lidov \citep{Kozai1962,Lidov1962} cycles that deliver planetesimals to the inner disc. 
In the case of a migrating gas giant, the dynamics of this process are quite complex.   In our simulations we find that the migrating planet efficiently scatters planetesimals that are not accreted onto high eccentricity orbits, often with semi-major axes beyond the formation location of the gas giant. The possibility of these planetesimals still being accreted by the gas giant would depend on the configuration of the companions in the system. More than half of the remaining planetesimals in our simulations are however, left on tight, low-eccentricity orbits between the star and the hot gas giant, and these planetesimals are very unlikely to be accreted. If we instead consider {\it in situ} formation, we expect the planetesimals in the outer disc to be left on relatively dynamically cold orbits. In this case, efficient inward scattering could occur in a vein similar to water delivery on Earth \citep[see e.g.,][]{Clement2018}, although this is again dependent on the exact configuration of the outer system. It is clear that further investigations of late accretion and potential enrichment of the atmospheres of hot Jupiters is required and we hope to address this topic in future research.

Finally, we have also made several simplifications regarding accretion and chemistry. We do not account for oxygen accreted from rocks within the ice-line, and we do not consider primoridal enrichment from volatiles accreted as gas during the planetary formation. We also do not account for oxygen binding with other elements in the atmosphere which may alter the abundance of water. In addition, the enrichment of giant planet atmospheres could be a result of convective mixing.  For example, heavy elements from the deep interior can be transported to the atmosphere if the planet is fully convective and the core is being eroded \citep[e.g., ][]{2020A&A...638A.121M} or due to erosion of primordial composition gradients \citep[e.g.,][]{2018A&A...610L..14V}. If the composition of the heavy elements in the deep interior is rich in refractory material this might also lead to enhancement in refractories in comparison to water and other volatiles.  However, it should be noted, that even in the {\it in-situ} formation scenario the planetary core likely forms beyond the water ice-line \citep{Batygin2016} and therefore this explanation is less likely. Nevertheless, the topic of convective mixing in giant planets and the relation between the atmospheric composition and the composition of the deep interior must be investigated in greater detail \citep{Helled2021}.

\section{Conclusions \& Outlook}
We have presented the results of numerical models of giant planets migrating through protoplanetary discs populated by planetesimals to explore the phenomenon of hot gas giants having atmospheres highly enriched in refractory elements but lacking in water. Our main conclusions can be summarized as follows: 
\begin{itemize}
    \item disc driven migration universally produces dynamically-cold, hot gas giants with super-stellar abundances of refractory elements
    \item this same process leaves planets with slightly super-stellar, stellar, or slightly sub-stellar abundances of water
    \item this model naturally explains the trend of low-inclination planets with high refractory-to-volatile abundance ratios

    \item these same relative enrichments cannot be explained with {\it in situ} models of gas giant formation, which place the refractory elements in the core of the planet
\end{itemize}
therefore, we suggest that alkali metal rich gas giants formed exterior to the snowline and then underwent significant accretion to reach their current orbits. 

Our models make strong predictions about the composition of hot gas giants that undergo disc-driven migration. Planets that form beyond the water ice-line will in all cases have an overabundance of metals relative to their stars, but this overabundance will be more pronounced in the refractory elements than volatiles, with water in some cases being present only in stellar or sub-stellar quantities. These outcomes are consistent with present observations, though the error bars on these observations remain large. 

Fortunately the outlook for improving on such observations is excellent. The PLATO mission \citep{Rauer2014} is expected to discover a host of new transiting hot giant planets around bright stars, making them perfect for atmospheric retrieval, and will even be able to perform some atmospheric characterisation on its own \citep{Grenfell2020}. The Ariel mission will then allow hitherto unforeseen precision in atmospheric retrieval \citep{Tinetti2018}, placing far tighter constraints on models such as ours. The launch of JWST will also allow for the characterisation of hot Jupiter atmospheres in both temperature and composition at up to several hundred parsecs from Earth \citep{Barstow2015}, allowing for C/O mixing ratios to be constrained to within 0.2dex \citep{Greene2016} even potentially allowing the characterisation of clouds \citep{Powell2018}. The combination of these three upcoming missions should provide a host of data for a large number of hot gas giants, allowing us to further constrain how important each of the canonical hot Jupiter formation and migration pathways are.

\section*{Acknowledgements}
 The authors thank Luis Welbanks, Nikku Madhusudhan, Anna Boehle, Richard Alexander, Simon M{\"u}ller, Nunzio Mannino, David Marshall and Martin Leibacher for useful discussions, and the anonymous referee for comments that improved the manuscript. RH acknowledges support from SNSF grant \texttt{\detokenize{200020_188460}} and the National Centre for Competence in Research ‘PlanetS’ supported by SNSF.
 
\section*{Data availability}
The data underlying this article will be shared on reasonable request to the corresponding author.
\bibliographystyle{mnras}
\bibliography{example} 

\appendix


\bsp	
\label{lastpage}
\end{document}